# Active Atoms and Interstitials in Two-dimensional Colloidal Crystals

Kilian Dietrich,[1] Ivo Buttinoni,[1, 2, *] Giovanni Volpe,[3] and Lucio Isa[1, †]

[1]*Laboratory for Interfaces, Soft Matter and Assembly,*
*Department of Materials, ETH Zurich, Zurich, Switzerland.*
[2]*Physical and Theoretical Chemistry Laboratory, Department of Chemistry, Oxford, United Kingdom*
[3]*Department of Physics, University of Gothenburg, Göteborg, Sweden.*

(Dated: October 24, 2017)

We study experimentally and numerically the motion of a self-phoretic active particle in two-dimensional (2D) loosely-packed colloidal crystals at fluid interfaces. Two scenarios emerge depending on the interaction between the active particle and the lattice: the active particle either navigates throughout the crystal as an interstitial or is part of the lattice and behaves as an active atom. Active interstitials undergo a run-and-tumble motion, with the passive colloids of the crystal acting as tumbling sites. Instead, active atoms exhibit an intermittent motion, which stems from the interplay between the periodic potential landscape of the passive crystal and the particle's self-propulsion. Our results shed new light on the behaviour of dense active phases and constitute the first step towards the realization of non-close-packed crystalline phases with internal activity.

Synthetic active particles, including phoretic microswimmers powered by local chemical [1], thermal [2] and ionic gradients [3], interact with confinements differently than Brownian particles [4]. Walls and microchannels are for instance effective pathways along which self-propelling colloids can be guided due to the persistent nature of their motion [5–7]. Similarly, pillars (or large passive spheres) can steer or trap active particles as a result of short-range (e.g., hydrodynamic) interactions between active particles and obstacles [8, 9]. Self-propelling particles confined in semi-dilute colloidal suspensions have also unveiled a wealth of new physics: active colloids can exert an effective swim pressure [10, 11] that promotes local compression and melting of clusters [12] as well as the segregation of active and passive species [13]. Experiments aimed at studying the motion of active particles in even denser phases, e.g. crystals or glasses, are far more challenging because the presence of neighbouring (active or passive) particles affects the local gradient responsible for self-propulsion [14, 15]. This fact leads to a reduced activity and to a short-range effective adhesion, effectively turning the active colloids into sticky spheres [10, 16]. Consequently, experimental studies of self-propelling colloids in dense structures lag behind progress in numerical simulations [17, 18] and have been so far limited to dry granular suspensions of macroscopic particles [19].

In this Letter, we study the motion of active colloidal particles within loosely-packed crystalline monolayers formed by spreading passive (Brownian) charged colloids at a flat water-oil interface [20], which permits us to avoid the reduction of activity happening in close-packed samples. In fact, long-range repulsive electrostatic dipolar forces [21] give rise to 2D crystalline arrays with lattice spacings significantly larger than the particle diameter. Depending on the their orientation at the interface, self-propelling Janus colloids show a varying degree of electrostatic coupling to the lattice and can behave either as *active interstitials* or as *active atoms*. For weak coupling, active interstitials are able to approach the passive colloids and navigate in the crystal in a run-and-tumble fashion. For large coupling, active atoms are effectively part of the crystal and self-propel while keeping a large distance (of the order of the lattice constant) from their neighbours.

We perform our experiments at a flat water-oil interface (see also [22], Section 1). Fluorescent polystyrene particles (PS, Microparticles GmbH) with radius $R = 1.4$ $\mu$m are placed at the interface via injection of a 1:1 water-isopropyl alcohol spreading solution [23]. A small amount ($\approx 1$ %) of the particles have one hemisphere coated by a 2 nm layer of Pt, which acts as a catalyst and promotes the decomposition of $H_2O_2$ in the aqueous phase, thus leading to active motion where both translation and rotation occur within the two-dimensional $xy$-plane of the interface [23, 24]. After spreading, the uncoated, passive PS particles self-assemble at the interface into loosely-packed crystalline monolayers (Fig. 1, c to e) due to dipolar, pair-additive, repulsive forces with a potential $U_{\rm pp}/(k_{\rm B}T) = a_{\rm pp}r^{-3}$, which stems from asymmetric charge effects across the interface [20, 25]. The lattice constant $L$ is determined by the amount of spread particles. At a given center-to-center distance $r$, the dipole magnitude depends on the surface charge and the contact angle ($\theta_{\rm p} = 123° \pm 8°$ [23]) of the passive particles pairs [25].

Upon exposure to the $H_2O_2$-rich aqueous phase, the Pt-coated PS particles swim with a velocity $V$, which, for a given $H_2O_2$ concentration, crucially depends on the orientation of the Pt-coated hemisphere with respect to the interface plane. Two predominant configurations, distinguishable by the particle brightness in fluorescence microscopy, are found (Video S1 [22]): particles either have the Pt-cap mostly wetted by the oil phase (active interstitials, Fig. 1(a)) or by the water phase (active atoms, Fig. 1(b)) [23]. The former configuration is typi-cally 5–10 times faster than the latter [23]. These active particles show similar contact angles ($\theta_a = 104° \pm 10°$ [23]) as the passive ones but, owing to the presence of the uncharged Pt cap, their orientation strongly affects the interactions with the surrounding passive colloids. The dipolar interaction with the particles in the lattice will therefore have a form $U_{pa} = cU_{pp}$, where $c$ is an orientation-dependent coupling parameter. Active interstitials (Fig. 1(a)) are characterized by a weak coupling and navigate through the crystal (blue trajectory in Fig. 1(c)) without perturbing it, as shown by the trajectories of the neighbouring passive colloids (black). Instead, active atoms (Fig. 1(b)) interact strongly with the crystal and their active motion leads to localized structural response of the monolayer (red trajectory in Fig. 1(c)). In the following, we consistently use blue and red colours for active interstitials and active atoms, respectively.

Our understanding of the experiments is verified by Brownian dynamics simulations of the passive and active microspheres ($R = 1.4\mu m$) [26]. The amplitude of the dipolar interactions $a_{pp} = 5 \cdot 10^{-13}\,\mathrm{m}^3$ is chosen in agreement with previous literature [21]. After equilibration of the lattice, we provide one microparticle with a free swimming velocity $V_0$ and coupling $c$. $V_0$ is chosen to match the experimental free swimming speed, i.e. the mean velocity in the absence of the crystalline landscape [23], for the peroxide concentration employed in our experiments ([$H_2O_2$] = 3%). Simulations at low coupling ($c = 0.01$, Fig. 1(d) and Video S2 [22]) reproduce the dynamics of active interstitials, whereas simulations at high coupling ($c = 1$, Fig. 1(e) and Video S3 [22]) exhibit features that are consistent with the dynamics of active atoms (Fig. 1(b)). Quantitative agreement is also found for the velocity (Fig 1(f)) and nearest-neighbour distance distributions (Fig. 1(g)). Concerning the latter, direct contact is only achieved at low coupling (blue data in Fig. 1(g)).

We first focus on the behaviour of the active interstitials. Figures 2(a-f) show three experimental (2(a-c), Videos S4-6 [22]) and numerical (2(d-f)) trajectories of active interstitials inside 2D crystals with decreasing lattice constant ($L = 22R$, $16R$ and $10R$). In all cases, the passive monolayer acts as a pinball machine: the active interstitial swims along straight paths with large persistence [23, 24] but reorients abruptly when it is in the proximity of a passive colloid. In fact, when the microswimmer approaches a passive particle (Fig. 2(g) and Video S7 [22]), its direction of motion (blue arrow) changes from the one observed before the collision and no longer corresponds to the 2D projection of the particle asymmetry (green arrow, note that in 2(g) a 20 nm Pt-layer is used to visualize the cap and this choice leads to a metastable tilted orientation [27]). Eventually, the microswimmer escapes the passive obstacle in a random direction (inset to the bottom-right panel in Fig. 2(g)). This scattering event randomizes the active-motion direc-

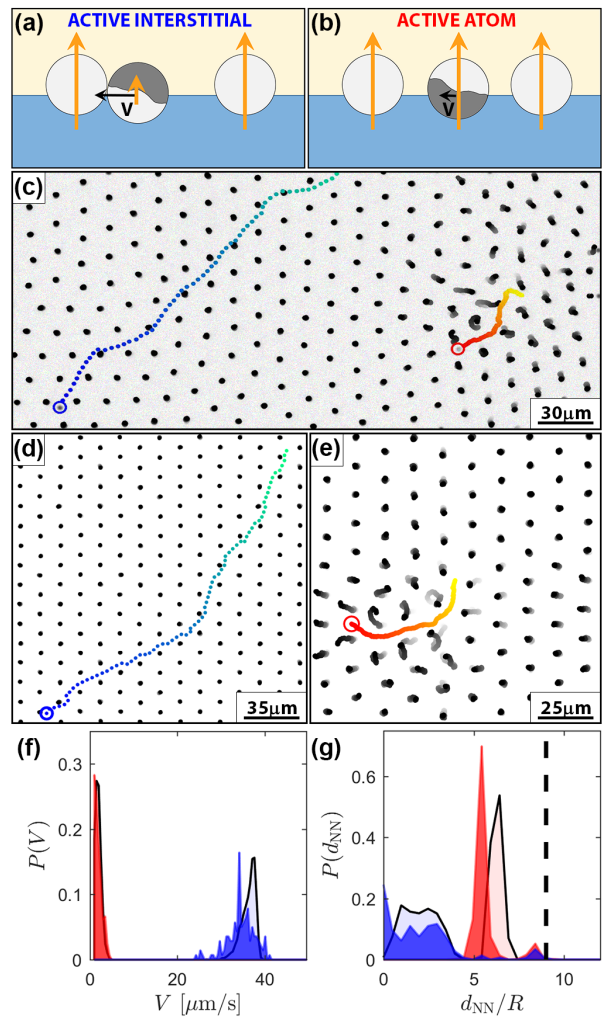

FIG. 1. Active particles in a colloidal crystal. (a,b) Interaction between active and passive particles at the water-oil interface: (a) active interstitial and (b) active atom. The orange arrows sketch the magnitude of the dipolar coupling. (c) Experimental trajectories and (d,e) corresponding numerical simulations. Blue: active interstitial. Red: active atom. Black: passive colloids. The colour gradients represent time from $t = 0$ s (green/yellow/gray) to $t = 30$ s (blue/red/black). The blue trajectories leave the field of view after 8 s. $L = 10R$, $V_0 = 38$ $\mu$m/s (active interstitials, coupling $c = 0.01$) and $V_0 = 6$ $\mu$m/s (active atoms, coupling $c = 1$). (f) Velocity and (g) normalized nearest-neighbour distance (surface-to-surface) probability distributions for the trajectories shown in (c), (d) and (e). Solid histograms: experiments. Transparent histogram: simulations. The dashed line in (g) marks the lattice constant. Experiments are made at [$H_2O_2$]=3% and images are recorded at 10 fps.

tion by enforcing a reorientation of the velocity vector. Orbiting of active particles around large circular obstacles has been observed in other recent works [8, 9] and has been ascribed to hydrodynamic interactions or to local deformations of the chemical gradient responsible for self-propulsion. In our system, there is another possible ness and decays as $r^{-4}$ (thus steeper than the dipolar repulsion) [28]. Assuming a surface roughness of the order of few nm and a coupling $c = 0.001$, the resulting potential (sum of dipolar repulsions and capillary attractions) close to contact is $\sim 50 k_B T$ (see [22], Section 4). These values fall within the range of effective temperatures [29] of active colloids reported by Ginot *et al.* [10], *i.e.* the self-propulsion is sufficiently strong to allow the particles to escape contact. To account for this phenomenon, in the simulations we have added an effective torque that leads to a reorientation of the active particle and acts only at short range, *i.e.* when the surface-to-surface distance is smaller than the particle diameter. The resulting motion of the active interstitial is reminiscent of the run-and-tumble trajectories of bacterial suspensions and the frequency of tumbles is determined by the probability of collisions with a passive colloid, as defined by the lattice constant of the crystal (Fig. 2(g)). At fixed [$H_2O_2$], the swimming velocity, evaluated as the maximum speed in the lattice (inset to 2(g)), remains constant and equal to the propulsion velocity measured at pristine interfaces

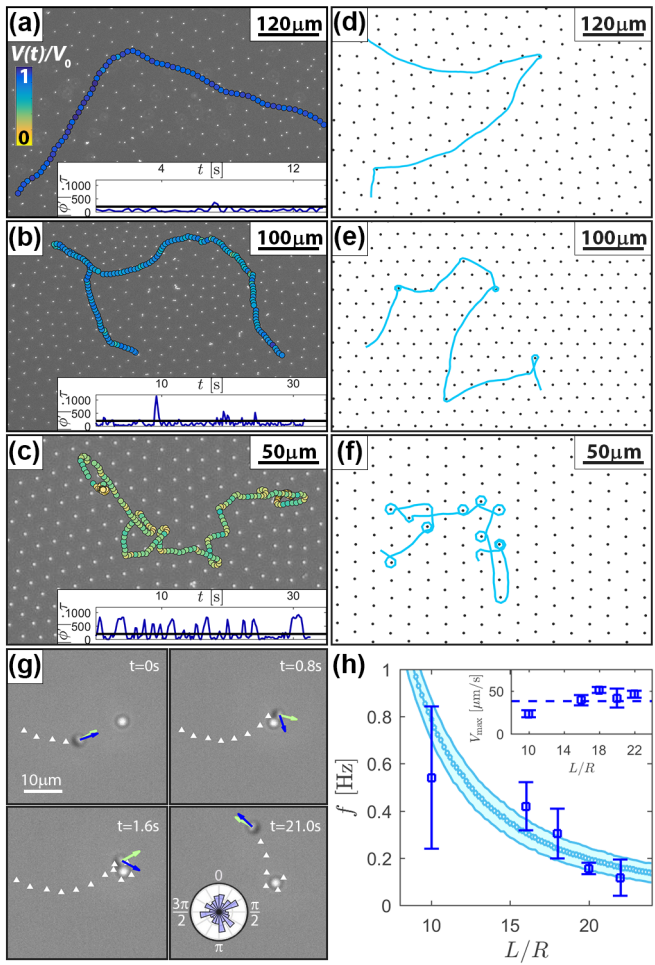

FIG. 2. Active interstitials. (a-c) Experimental trajectories of active interstitials in colloidal lattices with (a) $L = 22R$, (b) $L = 16R$ and (c) $L = 10R$ ([$H_2O_2$] = 3%, $V_0 = 38$ $\mu$m/s). The insets show the experimental absolute angular velocity $|\phi'|$ measured in sequential frames. When $|\phi'|$ reaches a value greater than $200\tau^{-1}$ (horizontal black line, $\tau$ is the free rotational diffusion time), a tumbling event is registered (see also [22], Section 2). The colour code highlights the relative particle speed. (d-f) Corresponding numerical simulations obtained using $c = 0.001$ and $V_0 = 38$ $\mu$m/s. (g) Frame sequence of an active interstitial interacting with a passive colloids. The direction of the swimming velocity (blue arrow) with respect to the particles orientation (green arrow) changes when the swimmer is captured. Details of the detection of the particle orientation are in [22], Section 3. Inset to the bottom-right panel: reorientation angle (i.e., difference between incident and escape angle) distribution for $\approx 200$ events. (h) Experimental (blue) and numerical (cyan) tumbling frequency plotted as function of the lattice constant. The inset shows that the maximum speed corresponds to the free swimming velocity $V_0 = 38$ $\mu$m/s (dashed horizontal line).

source of transient trapping. Because of contact line undulations, two particles confined at a fluid-fluid interface experience reciprocal capillary attractions whose magnitude depends on their contact angle and surface rough-

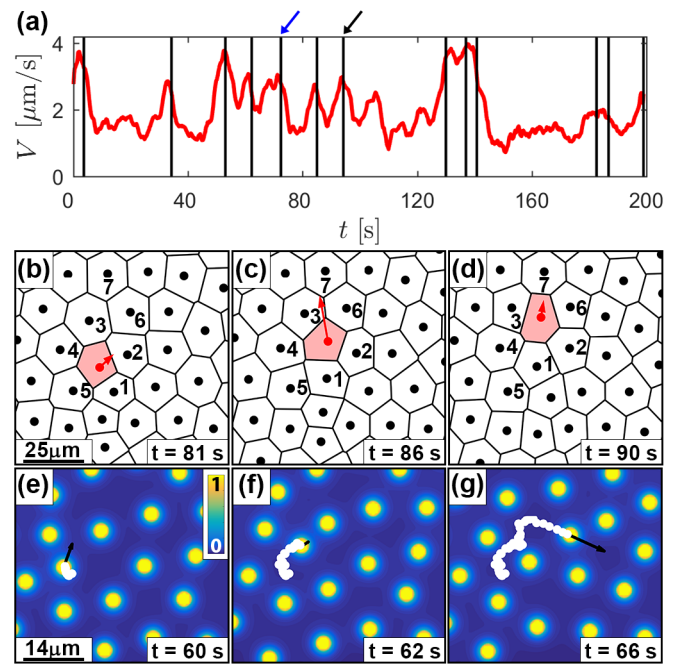

FIG. 3. Active atoms. (a) Time evolution of the velocity of an active atom in a crystalline monolayer with $L = 10R$ ([$H_2O_2$] = 3%, $V_0 = 6$ $\mu$m/s). The black vertical lines mark the times at which two (or more) nearest neighbours of the active particle change, suggesting a cell swap. (b-d) Corresponding Voronoi diagram (b) before, (c) during and (d) after the cell swap indicated by the black arrow in (a). (e-g) Potential landscape and trajectory of the active atom (e) before, (f) during and (g) after the cell-swap indicated by the blue arrow in (a). The potential is normalized by the value at the particle surface. In (b-g) the length of the arrows is proportional to the particle velocity.

(dashed line). A small decrease is only observed for the densest crystal ($L = 10R$).

As opposed to active interstitials, active atoms swim at distances from the passive particles that are significantly larger than $2R$ (Fig. 1(g), red data) and do not experience capillary trapping. Due to strong coupling ($c \approx 1$), active atoms feel the local electrostatic potential landscape of the two-dimensional crystal and exhibit large velocity fluctuations as they explore several unit cells (Fig. 3(a)). In particular, events in which the active atom swaps cell are marked by the vertical lines and correspond to the velocity peaks. Figures 3(b-d) show the Voronoi tessellation of an image sequence taken across one of the peaks in Fig. 3(a). The velocity of the active atom, proportional to the length of the red arrow, increases abruptly (Fig. 3(c) and Video S8 [22]) when it points towards one of the Voronoi vertices, leading to a sudden cell swap (Fig. 3(d)). A further description of the cell-swap mechanism is in [22], Section 5. At a first glance, this scenario can be compared to driving a colloidal particle at constant force through a 2D periodic potential (e.g., a laser pattern) [30], where the velocity is affected by its direction relative to the symmetry axes and by its local distance from the potential minima. Here, however, the active particle contributes itself to the energy landscape by deforming it continuously while it swims through the lattice. The dynamics is therefore better understood by plotting the instantaneous local surrounding potential, which is depicted for another cell swap in Figure 3(e-g): in agreement with the Voronoi description, the swimming velocity shows maxima when the active atom self-propels through local potential minima (Fig. 3(f) and Video S9 [22]).

Experimentally, we are unable to know *a priori* the exact value of the coupling between each active atom and the lattice passive particles. The contact angle of Pt-coated particles confined at water-oil interfaces can in fact vary up to 20° [23], which implies uncertainties on the coupling $c$ (see [22], Section 6). We therefore use numerical simulations to address the full dynamical response as a function of the control parameters, $L$, $V_0$ and $c$. We compute the mean coordination number $N_{\text{coord}}$ (Fig. 4(a)) and the mean residual activity $V/V_0$ (Fig. 4(b)) of active particles that self-propel through colloidal crystalline monolayers for 500 s. The former reveals the extent to which self-propelling colloids belong to the hexagonal crystals and reaches its maximum value ($N_{\text{coord}} \approx 6$) when highly-coupled ($c \to 1$) and slow ($V_0 \to 0$) particles roam in dense lattices (bottom stack in Fig. 4(a)). Larger swimming velocities locally destroy the hexagonal structure, whereas at lower couplings the active particles interact less with the lattice. The interplay between the active motion and the crystalline environments becomes weaker in sparser monolayers (top stack in Fig. 4(a)). Similarly, a large loss of activity is reported in dense lattices and for high couplings (top stack in Fig. 4(b)), whereas active interstitials (bottom stack in Fig. 4(b), $c \to 0$) slow down only upon tumbling. A rough estimation of the experimental $c$ can be done by measuring the mean distance of the active particle from the nearest neighbour $\langle d_{NN} \rangle$, at fixed $L$ and $V_0$ (see [22], Section 6). Figure 4(c) shows numerical (lines) and experimental (symbols) data of the residual activity for $L = 10R$: in both instances, $V/V_0$ decreases with $c$ as a result of the stronger interaction with the crystalline landscape, although the loss is less pronounced in the numerical simulations because hydrodynamic and capillary interactions near contact are neglected.

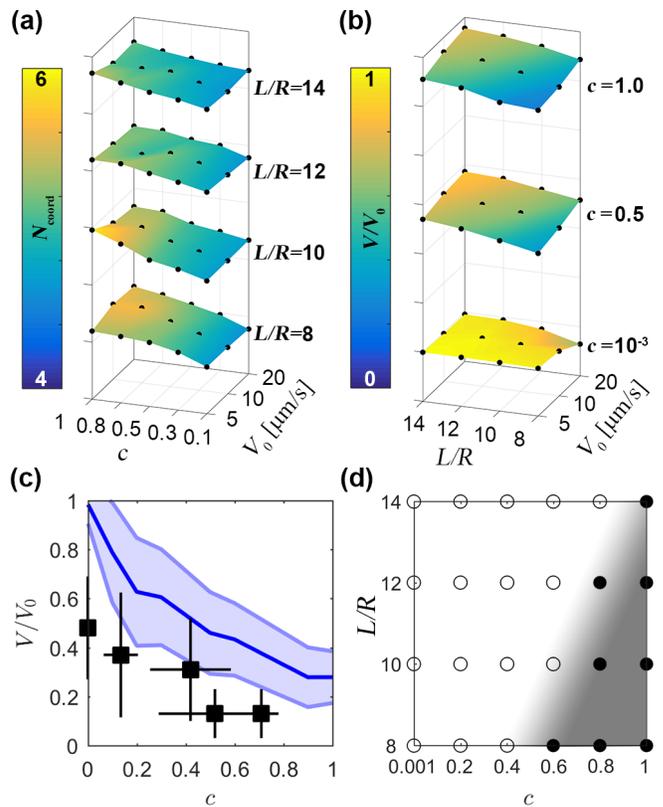

FIG. 4. (a,b) Numerical colour maps of (a) the coordination number $N_{\text{coord}}$ and (b) residual activity $V/V_0$, *i.e.* the mean velocity $V$ normalized by the mean free swimming speed $V_0$, as a function of $V_0$, $L$ and $c$. The dots correspond to the simulated values. (c) Experimental (black symbols) and numerical (blue line) mean velocity normalized by the maximum speed in the lattice $V_{\text{max}}$ ($\approx V_0$), plotted as a function of the coupling $c$. The blue band marks the standard deviation of the numerical results. (d) Numerical data for fixed free swimming velocity ($V_0 = 1 \ \mu\text{m/s}$) and variable $L$ and $c$, revealing mobile (empty symbols) and caged (filled symbols) states. A particle is labelled as caged if the maximum value of its root mean square displacement is smaller than $L$. The grey band highlights the transition.

The results presented in Figs. 4(a-c) are valid provided that the active particle is able to explore the surrounding environment. At small $V_0$ and large $c$ the active particle might be caged in one lattice position (solid circles in Fig. 4(d), Video S10 [22]), while remaining free to move for smaller coupling values (empty circles), or larger $V_0$. Remarkably, a zero-order approximation of the swimming velocity required to avoid caging can be obtained by comparing Stokes' drag $F_s \sim 6\pi\eta R V_0$ to the force needed to overcome the potential barrier $F_p \sim \nabla U/L$, as shown in [22], Section 7.

In conclusion, we have investigated the motion of active particles in 2D loosely-packed crystals at a fluid interface as a function of their coupling with the surrounding lattice. We identified two types of swimmers: *active interstitials*, which have a weak coupling to the structure and move in a run-and-tumble fashion, and *active atoms*, which interact with the crystalline potential landscape by deforming the lattice. Our results showcase the potential of fluid interfaces to realize new active phases, en route toward using active atoms as the building-blocks for active materials.

We thank Paolo Malgaretti for fruitful discussions. KD, IB and LI acknowledge financial support from the Swiss National Science Foundation grant PP00P2 144646/1 and the ETH Zurich research grants ETH-16 15-1 and FEL-02 14-1. GV acknowledges funding by the European Commission with ERC Starting Grant ComplexSwimmers (grant number 677511).

* ivo.buttinoni@chem.ox.ac.uk
† lucio.isa@mat.ethz.ch

[1] J. R. Howse, R. A. Jones, A. J. Ryan, T. Gough, R. Vafabakhsh, and R. Golestanian, Phys. Rev. Lett. **99**, 048102 (2007).
[2] H.-R. Jiang, N. Yoshinaga, and M. Sano, Phys. Rev. Lett. **105**, 268302 (2010).
[3] J. Zhang, J. Yan, and S. Granick, Ang. Chemie Int. Ed. **55**, 5166 (2016).
[4] C. Bechinger, R. Di Leonardo, H. Löwen, C. Reichhardt, G. Volpe, and G. Volpe, Rev. Mod. Phys. **88**, 045006 (2016).
[5] S. Das, A. Garg, A. I. Campbell, J. Howse, A. Sen, D. Velegol, R. Golestanian, and S. J. Ebbens, Nat. Commun. **6** (2015).
[6] J. Simmchen, J. Katuri, W. E. Uspal, M. N. Popescu, M. Tasinkevych, and S. Sánchez, Nat. Commun. **7**, 10598 (2016).
[7] A. Mozaffari, N. Sharifi-Mood, J. Koplik, and C. Maldarelli, Phys. Fluids **28**, 053107 (2016).
[8] D. Takagi, J. Palacci, A. B. Braunschweig, M. J. Shelley, and J. Zhang, Soft Matter **10**, 1784 (2014).
[9] A. T. Brown, I. D. Vladescu, A. Dawson, T. Vissers, J. Schwarz-Linek, J. S. Lintuvuori, and W. C. Poon, Soft Matter **12**, 131 (2016).
[10] F. Ginot, I. Theurkauff, D. Levis, C. Ybert, L. Bocquet, L. Berthier, and C. Cottin-Bizonne, Phys. Rev. X **5**, 011004 (2015).
[11] S. C. Takatori, R. De Dier, J. Vermant, and J. F. Brady, Nat. Commun. **7** (2016).
[12] F. Kümmel, P. Shabestari, C. Lozano, G. Volpe, and C. Bechinger, Soft Matter **11**, 6187 (2015).
[13] J. Stenhammar, R. Wittkowski, D. Marenduzzo, and M. E. Cates, Phys. Rev. Lett. **114**, 018301 (2015).
[14] I. Buttinoni, J. Bialké, F. Kümmel, H. Löwen, C. Bechinger, and T. Speck, Phys. Rev. Lett. **110**, 238301 (2013).
[15] J. Palacci, S. Sacanna, A. P. Steinberg, D. J. Pine, and P. M. Chaikin, Science **339**, 936 (2013).
[16] I. Theurkauff, C. Cottin-Bizonne, J. Palacci, C. Ybert, and L. Bocquet, Phys. Rev. Lett. **108**, 268303 (2012).
[17] B. van der Meer, L. Filion, and M. Dijkstra, Soft Matter **12**, 3406 (2016).
[18] C. A. Weber, C. Bock, and E. Frey, Phys. Rev. Lett. **112**, 168301 (2014).
[19] G. Briand and O. Dauchot, Phys. Rev. Lett. **117**, 098004 (2016).
[20] P. Pieranski, Phys. Rev. Lett. **45**, 569 (1980).
[21] B. J. Park, B. Lee, and T. Yu, Soft Matter **10**, 9675 (2014).
[22] Supplementary Material available at.
[23] K. Dietrich, D. Renggli, M. Zanini, G. Volpe, I. Buttinoni, and L. Isa, New J. Phys. **19**, 065008 (2017).
[24] X. Wang, M. In, C. Blanc, M. Nobili, and A. Stocco, Soft Matter **11**, 7376 (2015).
[25] K. D. Danov and P. A. Kralchevsky, J. Colloid Interf. Sci. **298**, 213 (2006).
[26] G. Volpe, S. Gigan, and G. Volpe, Am. J. Phys. **82**, 659 (2014).
[27] X. Wang, M. In, C. Blanc, P. Malgaretti, M. Nobili, and A. Stocco, Faraday Diss. **191**, 305 (2016).
[28] K. D. Danov, P. A. Kralchevsky, B. N. Naydenov, and G. Brenn, J. Colloid Interf. Sci. **287**, 121 (2005).
[29] M. Han, J. Yan, S. Granick, and E. Luijten, Proc. Natl. Acad. Sci. **114**, 7513 (2017).
[30] P. T. Korda, M. B. Taylor, and D. G. Grier, Phys. Rev. Lett. **89**, 128301 (2002).

# Active Atoms and Interstitials in Two-dimensional Colloidal Crystals: Supplementary Material

Kilian Dietrich,[1] Ivo Buttinoni,[1,2] Giovanni Volpe,[3] and Lucio Isa[1]

[1]*Laboratory for Interfaces, Soft Matter and Assembly, Department of Materials, ETH Zurich, Zurich, Switzerland.*
[2]*Physical and Theoretical Chemistry Laboratory, Department of Chemistry, Oxford, United Kingdom*
[3]*Department of Physics, University of Gothenburg, Göteborg, Sweden.*

(Dated: October 24, 2017)

## 1. FABRICATION OF ACTIVE PARTICLES

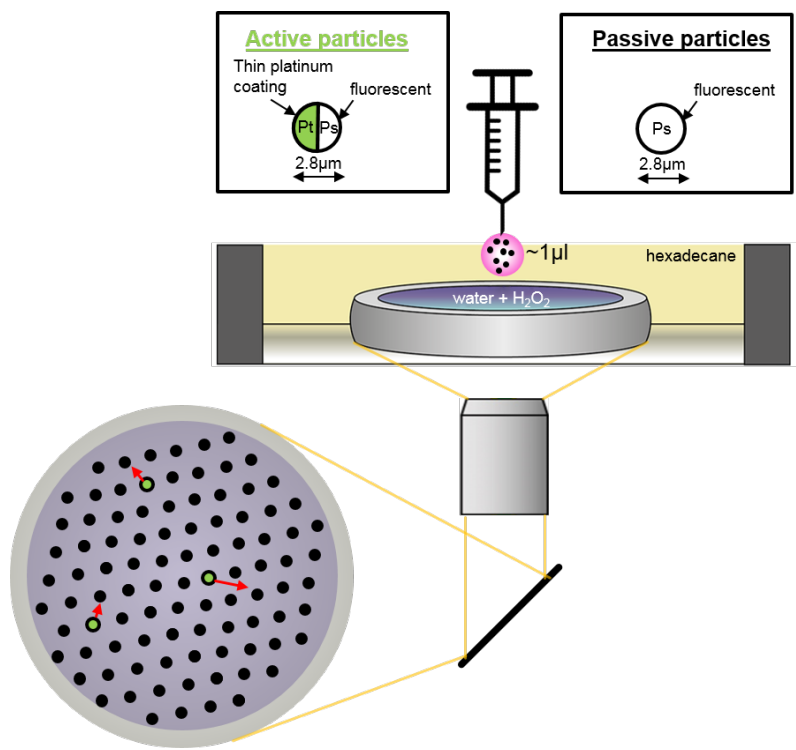

FIG. S1. Particle deposition onto a flat water/hexadecane interface. The spreading solution (isopropyl alcohol + water, pink) contains active (coated, green) and passive (uncoated, black) PS particles. After the deposition, dipolar interactions cause the formation of two-dimensional loosely-packed colloidal crystals containing also active particles. The orientation of the active particle with respect to the interface makes it behave either as an interstitial or as an active atom.

We assemble monolayers of sulphate PS particles (radius R = 1.4 $\mu$m, $SO_4^-$ functional groups, Microparticles GmbH) by drying a 200 $\mu$l droplet of a diluted suspension (35 % w/v) on a pre-cleaned microscope slide (Thermo Scientific). We then sputter 2 nm of platinum onto the monolayer with a 90-degrees glancing angle in order to coat one hemisphere of the particles. The process pressure is set to 0.5 Pa to obtain a smooth deposition. The particles are released from the substrate into 50 ml milliQ water via ultra-sonication for approximately 2 minutes.

Non-stabilized hydrogen peroxide ($H_2O_2$, 30 % w/v, Merck Millipore) is first added to milliQ water to prepare fuel-rich aqueous solutions of given concentrations. The oil phase consists of hexadecane (Arcos Organics) that is preventively purified through a column containing alumina powder (EcoChromTM, MP Alumina B Act.1) and silica gel 60 (Merck) in order to remove surface-active contaminants. A customized measuring cell (Fig. S1) is made of two rings of different heights: an inner stainless steel ring (5 mm inner diameter) and an outer Teflon ring (2 cm inner diameter). The inner ring is filled with the water/$H_2O_2$ solution until the surface is pinned to the edge. Hexadecane is then poured in the rest of the cell in order to create a flat oil/water interface as shown in Fig. S1. Active particles and uncoated PS particles are dispersed in a 50:50 water/isopropyl alcohol spreading solution and sonicated for approximately 10 minutes. Finally, 1 $\mu$l of spreading solution is placed at the liquid-liquid interface to create two-dimensional loosely-packed colloidal crystals (Fig. S1) using a thin pre-cleaned syringe (Exmire Microsyringe cleaned with ethanol and deionized water). The different lattice spacings are achieved by tuning the concentration of uncoated particles in the spreading solution (0.6 % w/v to 0.1 % w/v).

The particles are imaged using an inverted optical microscope operated in fluorescence mode (Zeiss, Axio Observer, 89north, PhotoFluor II). Videos are recorded with a high resolution camera (2560 x 2160 pixels, 12 bits) at 5 – 10 fps for up to 600 seconds and analysed with custom Matlab codes.

## 2. DETECTION OF TUMBLING EVENTS

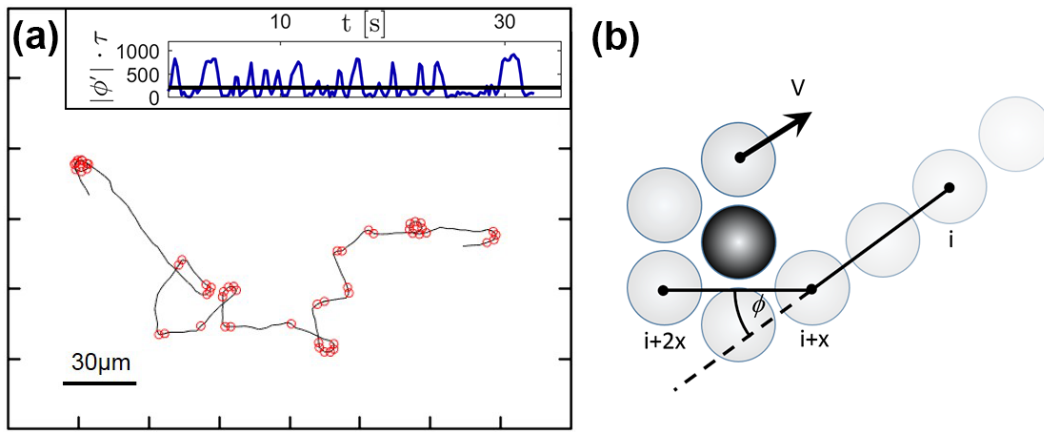

FIG. S2. (a) Trajectory of an active interstitial (black line). Tumbling events are marked using red open circles. Inset: absolute angular velocity normalized by the rotational diffusivity $D_\mathrm{R} = 1/\tau$. If this quantity is larger than $200\tau^{-1}$ a tumbling event is registered. (b) Sketch of a single tumbling. An active interstitial (grey) is reoriented by a passive particle (black).

Tumbling events are determined by monitoring the absolute angular velocity $|\phi'|$ of the active particle (interstitial) in consecutive frames ($\Delta t = 0.4$ s). In case of tumbling, the particle reorients much faster than a freely swimming particle. We arbitrarily choose a threshold of $200\tau^{-1}$ (roughly corresponding to a 30° change of propulsion direction) to register a tumbling event (Figure S2(a) and inset). We also apply two additional criteria to identify a tumbling event: i) the passive colloid has to be within $6R$ from the active interstitial (to exclude events corresponding to spontaneous reorientation) and ii) only two-particle interactions are considered. Figure S2(b) shows how the angle $\phi$ at the position $\mathbf{r}_{i+x}$ is extracted by comparing the vectors $\mathbf{r}_{i+x} - \mathbf{r}_i$ and $\mathbf{r}_{i+2x} - \mathbf{r}_{i+x}$.

## 3. ORIENTATION OF ACTIVE INTERSTITIALS

In Fig. 2(g) of the main text, we use particles with a think Pt coating (20 nm) in order to study the physics behind a tumbling event. In order to track the centres of mass of the coated and bare hemispheres, we apply a dual-channel illumination where a metal halide lamp (89 North) and bright-field illumination are simultaneously used. Connecting the centres of mass gives the vector of orientation for one particle in the respective frame. In Figure 2(g), we also show the velocity vector extracted from the spatial displacement of the particle in sequential frames. The velocity and orientation vectors are aligned when the particle is freely moving. Misalignments occur upon tumbling, indicating that passive particles enforce a reorientation of the velocity of active interstitials.

We remark that particles with thick coatings ($> 5$ nm) do not uptake the two main configurations illustrated in the main manuscript and Ref. [1]. In this case, the particle orientation is determined by surface roughness, i.e. by pinning of the three-phase contact line, rather than by the wetting properties of the two hemispheres. All the other experiments shown in the main manuscript are performed using 2 nm Pt-coatings.

## 4. CAPILLARY ATTRACTIONS AND ELECTROSTATIC REPULSIONS

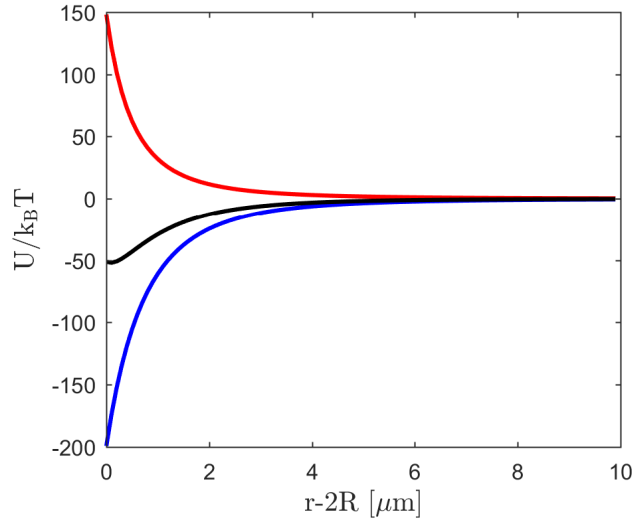

FIG. S3. Estimation of the potentials experienced by an active interstitial in a 2D crystal of passive particles, plotted as a function of the surface-to-surface distance $r - 2R$. (red) Dipolar electrostatic repulsion. (blue) Capillary attraction. (black) Total potential.

In the main manuscript, we suggest that the orbiting of active interstitials around the passive particles of the lattice might stem from the interplay between self-propulsion and the total potential acting on the active colloids near contact. Here, we quantify this potential, which is the sum of a *capillary attraction* and a *dipolar electrostatic repulsion*.

*Dipolar repulsion*: The dipolar repulsive potential (Fig. S3, red curve) is due to charge asymmetries across the interface [2–4] and has the following form:

$$\frac{U_{\text{dipole}}}{k_{\text{B}}T} = c\frac{a_{\text{pp}}}{r^3}, \quad (1)$$

where $a_{\text{pp}} = 5 \cdot 10^{-13}$ $m^3$ is the dipolar amplitude and $c = 0.001$ is the value of coupling that faithfully mimics the motion of active interstitials (Figs. 2(a-c) of the main paper).

*Capillary attraction*: The capillary attraction is quadrupolar and is due to contact line undulations triggered by the particle surface roughness [5]. The corresponding potential (Fig. S3, blue) is:

$$\frac{U_{\text{capillary}}}{k_{\text{B}}T} = -\frac{12\pi\gamma H_{\text{a}} H_{\text{p}} R^4}{r^4}, \quad (2)$$

where $\gamma$ is the oil-water surface tension, and $H_{\text{a}}$ and $H_{\text{p}}$ correspond to the surface roughness of the active interstitial and the passive particle, respectively.

The sum of these two contributions (black curve) leads to an effective potential with a minimum of the order of few tens of $k_{\text{B}}T$ close to contact but which is not strong enough to induce permanent adhesion. In fact, active particles have an effective thermal energy in the range of 10 to $10^3 k_{\text{B}}T$ [6] which allows them to occasionally detach.

In the following, we bring additional evidence that the afore-mentioned balance between capillary attraction and dipolar electrostatic repulsion may be at the origin of the observed behaviour. In order to do so, we modify our sample preparation protocol to disable the electrostatic interaction in our system. To this aim, we run the preparation process for active particles as described above but redisperse the coated particles into pure isopropyl alcohol instead of water. Without water, the functional head-groups of our particles cannot deprotonate, which leaves them uncharged. Spreading of such particles onto a water/hexadecane interfaces then leads to active swimmers that agglomerate regardless of their orientation relative to the interface (Fig. S4(a-c), fuel is present in the water phase). Figure S4(d) shows a larger field of view of the same spot in the cell after 5 minutes.

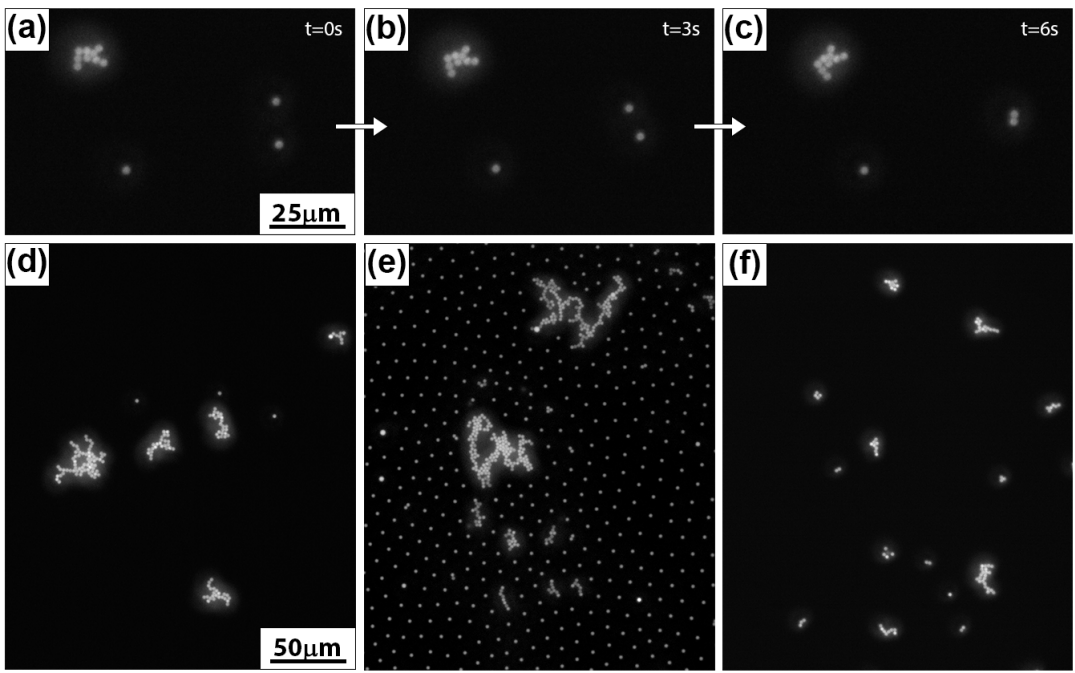

FIG. S4. (a-c) Capillary attraction leads to agglomeration of active swimmers (electrostatics repulsion is disabled). (d) Larger field of view of the same cell as in (a-c) after 5 minutes. (e) Consecutive spreading of passive particles (regular spreading solution) and coated particles (isopropyl alcohol as spreading solution). (f) Both coated and uncoated particles spread using isopropyl alcohol only.

A similar effect can be observed when the PS particles are first spread with a regular 50:50 water/isopropyl alcohol spreading solution, which allows for the lattice formation, and then the coated particles are spread from pure isopropyl alcohol, as demonstrated in Figure S4(e). Here, in the absence of any hydrogen peroxide, a large portion of the passive lattice remains stable but agglomerates are formed by coated particles, indicating that aggregation is not driven by chemical gradients. Finally, if both coated and uncoated particles are first dried and then redispersed in pure isopropyl alcohol, only aggregates of these particles are seen at the interface after spreading (Figure S4(f)). These observations confirm the necessity of electrostatic effects to form the passive lattices and to ensure strong coupling between active and passive particles. They moreover emphasize that attractive capillary forces exist in our system.

## 5. CELL SWAP

Active atoms undergo an intermittent motion and deform (locally) the lattice structure. In Fig. 3a of the main article, we show that the particle velocity shows a peak during a cell swap. Here, we look at the details of this mechanism. Figure S5 shows experimental (Fig. S5(a)) and numerical (Fig. S5(b)) data of one active atom swapping multiple cells during its motion. $d/R$ indicates the distance between the local position of the particle and the segment connecting the two nearest neighbours. Negative and positive values correspond to the active atom swimming towards and away the cell edge, respectively.

In addition to an overall velocity increase at $d/R \approx 0$, consistent with Fig. 3a of the main manuscript, $V/V_{\max}$ drops shortly before swapping cell and increases shortly after. This fluctuation stems from the particle moving up and down the local potential barrier, respectively at $d/R < 0$ and $d/R > 0$.

## 6. ESTIMATION OF COUPLING PARAMETERS

The dipolar electrostatic potential between two passive particles is $U_{\mathrm{pp}}/(k_\mathrm{B} T) = a_{\mathrm{pp}} r^{-3}$, where the strength of the interactions $a_{\mathrm{pp}} \propto p_\mathrm{p}^2$ and $p_\mathrm{p}$ is the dipole moment of a passive particle. For polystyrene colloids, the interactions are expected to be mediated through the oil phase [3]. In this case,

$$p_\mathrm{p} = 4\pi\sigma_\mathrm{p} D(\theta_\mathrm{p}, \epsilon_\mathrm{pn}) R^3 sin^3(\theta_\mathrm{p}),  \quad (3)$$

where $\sigma_\mathrm{p}$ is the surface charge density, $\theta_\mathrm{p}$ is the particle contact angle, $\epsilon_\mathrm{pn}$ is the ratio between the dielectric constants of the particle and the oil ($\approx 1.3$ for polystyrene/n-hexadecane) and $D(\theta_p, \epsilon_{pn})$ is a dimensionless function that was calculated and tabulated in Ref. [7].

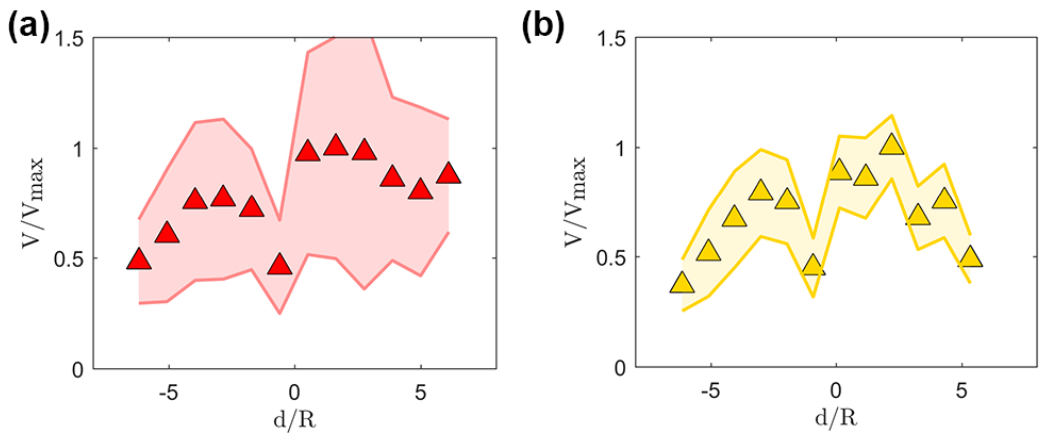

FIG. S5. (a) Experimental and (b) numerical data of an active atom swapping cells in a loosely-packed lattice. $V_0 \approx 5$ $\mu$m/s, $L/R = 10$, $c = 0.4$. The velocity has been smoothed over 2s and data binning was done according to the Freedman-Diaconis rule.

Given our definition of the coupling parameter as $c = U_\mathrm{pa}/U_\mathrm{pp}$ and the above expression for the interaction potential and the dipole moment of a passive particle, the definition of $c$ reduces to

$$c = \frac{p_\mathrm{a}}{p_\mathrm{p}} = \frac{\sigma_\mathrm{a} D(\theta_\mathrm{a}, \epsilon_\mathrm{pn}) sin^3(\theta_\mathrm{a})}{\sigma_\mathrm{p} D(\theta_\mathrm{p}, \epsilon_\mathrm{pn}) sin^3(\theta_\mathrm{p})}. \quad (4)$$

For the case of the active atoms, which have the Pt cap pointing mostly into the water, we can assume that $\sigma_\mathrm{a} \approx \sigma_\mathrm{p}$, since both passive and active particle have the polystyrene surface exposed to the oil. Therefore, the possible values of the coupling parameter are defined by the relative difference in contact angles between passive and active particles. Consequently, the coupling parameter will range between $c = 1$ when the two particles have the same contact angles to a lower bound of $c \approx 0.4$ for the biggest difference in contact angles using the values reported in the main text and previously measured in Ref [1].

Experimentally, we do not know *a priori* to what extent an observed active atom couples to the potential landscape. We estimate this parameter by looking at the nearest-neighbour distance $d_\mathrm{NN}$, averaged over the entire trajectory. The experimental value of $\langle d_\mathrm{NN} \rangle$ is compared to numerical data at different $c$ in order to extract the experimental coupling. Figure S6 shows the distribution of $c$ for active atoms at $V_0 = 0$ (no fuel in the water phase): the boundaries agree with the estimation provided above. In contrast, Figure S7 shows two extreme cases of low (Fig. S7(a)) and high (Fig. S7(b)) coupling for active atoms swimming at 6 $\mu$m/s. The presence of self-propulsion leads to an underestimation of the coupling parameter using the nearest-neighbour distance as a proxy, since activity allows the atoms to get closer to the passive particles. Nonetheless, we emphasize that this effect is present both in experiments and in numerical simulations, thus making the comparison between the two self-consistent.

For the case of the active interstitials, the fact that the Pt surface is exposed to the oil implies $\sigma_a \ll \sigma_p$, leading to up vanishingly small values of $c$.

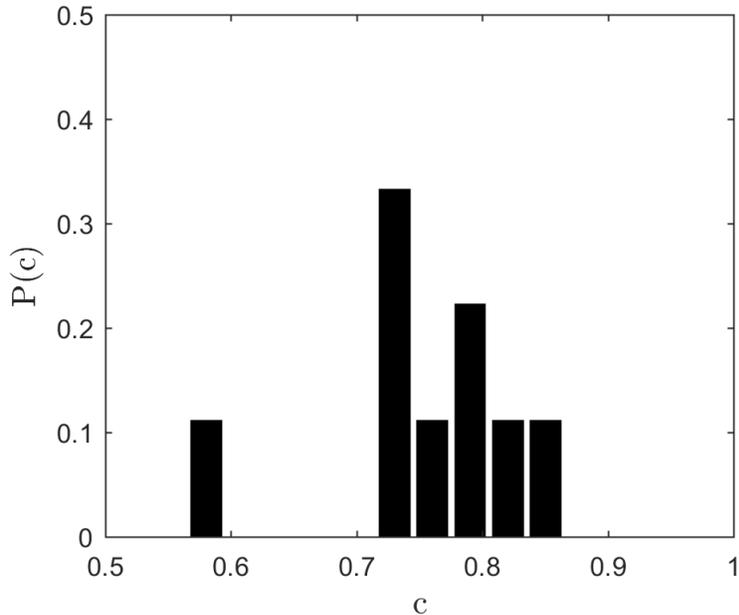

FIG. S6. Distribution of the coupling $c$ of N=9 atoms in the absence of self-propulsion. Values are in agreement with the estimates coming from our estimations of the dipolar interactions at fluid-fluid interfaces.

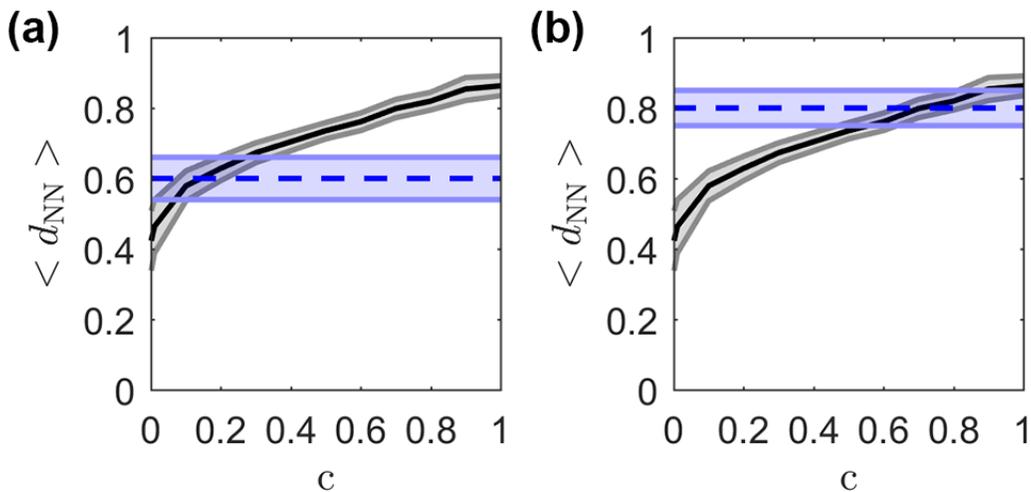

FIG. S7. Method to extract the coupling $c$ of experimental data. The particle in (a) couples less strongly to the potential landscape than the particle in (b). $c$ is extracted as the intercept between the experimental mean nearest neighbour distance $\langle d_{NN} \rangle$ (blue) and the numerical curve at various $c$ (black line).

## 7. CAGING

In the following, we roughly estimate the swimming velocity $V_{0,\min}$ required to avoid caging for an active particle of radius $R = 1.4$ $\mu$m in a loosely-packed crystal of passive spheres with lattice spacing $L$. In this simple model, caging is determined by the balance of two main forces: a *caging force* $F_\mathrm{p}$ related to the crystalline potential landscape, which holds the particle in the lattice cell, and the *propulsion force* $F_\mathrm{s}$, which promotes the escape.

*Caging force $F_\mathrm{p}$*: The dashed line in Fig. S8 identifies the path of minimum resistance, *i.e.* the direction at which the active particle (red) faces the smallest potential barrier $\nabla U = U_B - U_A$ due to neighbouring passive colloids (grey). In a very good approximation, for coupling $c = 1$ the crystalline potential in (A) and (B) point is:

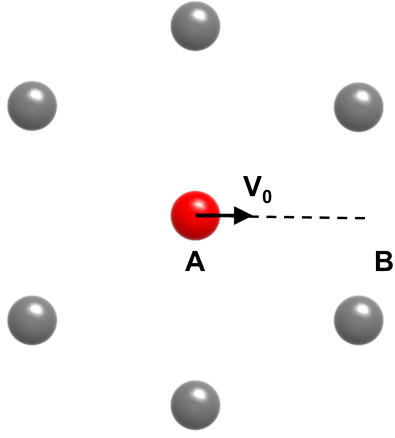

FIG. S8. Sketch of an active particle (red) with swimming velocity $V_0$ caged in a crystalline cell of passive particles (grey).

$$\frac{U_A}{k_B T} = 6 \frac{a_{pp}}{L^3}, \tag{5}$$

$$\frac{U_B}{k_B T} = 2 \frac{a_{pp}}{(L/2)^3} = 16 \frac{a_{pp}}{L^3}, \tag{6}$$

where $a_{pp}$ is the coupling amplitude. We then make the assumptions that the active atom moves between (A) and (B) at constant speed $V_0$ without deforming the crystalline cell. The caging potential force is:

$$F_p \approx \frac{U_B - U_A}{L} = 10 \frac{a_{pp}}{L^4} (k_B T). \tag{7}$$

*Propulsion force $F_p$*: Since the system is overdamped, the propulsion force is equal to the Stokes' drag, *i.e.*:

$$F_s = 6\pi \eta R V_0, \tag{8}$$

where $\eta = 1.7$ mPa·s is the average of the water and oil viscosity.

By equating $F_s$ to $F_p$, we can extract, for any value of $L$, the minimum swimming velocity $V_{0,\min}$ that allows the active particle to move out of the crystalline cell as shown in Fig. S8. The values of $V_{0,\min}$ are reported in Table I. In spite of the crude approximations (e.g., no lattice deformation and constant speed), this simple model remarkably leads to values of $V_{0,\min}$ whose order of magnitude is in agreement with Fig. 4d of the main manuscript.

TABLE I. Potential force and minimum escape swimming velocity for different lattice spacings.

| $L/R$ | $F_p$ [pN] | $V_{0,\min}$ [$\mu$m/s] |
|---|---|---|
| 10 | 0.3 | 11 |
| 15 | 0.06 | 2 |
| 20 | 0.02 | 0.74 |
| 30 | 0.004 | 0.14 |

Using numerical simulations, we calculate the escape velocity for different values of $c$ and $L$ (see Fig. 4(d) of the main manuscript). In our experiments, active atoms have a free swimming velocity of $\approx 6$ $\mu$m/s, with $\approx 2.5$ $\mu$m/s standard deviation. In Figure S9 we show simulation data for $V_0 = 5$ $\mu$m/s and demonstrate that no caging is expected in our system. Cases in which the particles are too slow to overcome the energy barrier are rare (e.g., Video S10).

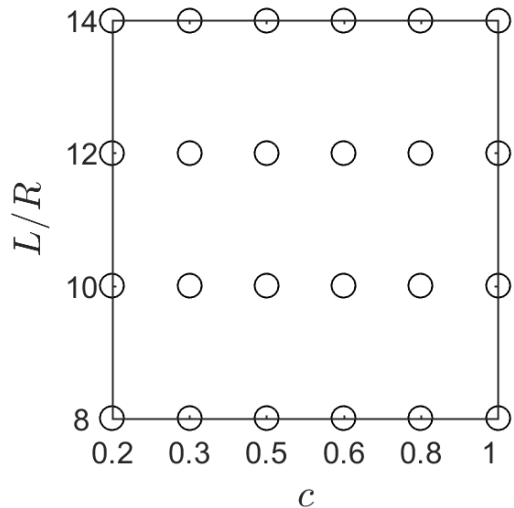

FIG. S9. Numerical simulations of caging of an active particle with free swimming velocity $V_0 = 5$ $\mu$m/s and coupling $c$ in colloidal crystals with lattice spacing $L$, in analogy to Fig. 4(d) in the main article. $V_0 = 5$ $\mu$m/s corresponds to the swimming velocity in our experimental system. Numerical simulations reveal no caging (open circles), in agreement with our experimental observations.

[1] K. Dietrich, D. Renggli, M. Zanini, G. Volpe, I. Buttinoni, and L. Isa, New J. Phys. **19**, 065008 (2017).
[2] P. Pieranski, Phys. Rev. Lett. **45**, 569 (1980).
[3] P. A. Kralchevsky, K. D. Danov, and P. V. Petkov, Phil. Trans. R. Soc. A **374**, 20150130 (2016).
[4] B. J. Park, B. Lee, and T. Yu, Soft Matter **10**, 9675 (2014).
[5] K. D. Danov, P. A. Kralchevsky, B. N. Naydenov, and G. Brenn, J. Colloid Interf. Sci. **287**, 121 (2005).
[6] F. Ginot, I. Theurkauff, D. Levis, C. Ybert, L. Bocquet, L. Berthier, and C. Cottin-Bizonne, Phys. Rev. X **5**, 011004 (2015).
[7] K. D. Danov and P. A. Kralchevsky, J. Colloid Interf. Sci. **298**, 213 (2006).